**Element-specific electronic and structural dynamics using transient X-ray spectroscopy**

Hanzhe Liu,[1] Isabel M. Klein,[1,*] Jonathan M. Michelsen,[1,*] Scott K. Cushing[1,†]

[1]Division of Chemistry and Chemical Engineering, California Institute of Technology, Pasadena, CA 91125, USA.

[*]These authors contributed equally.

[†]Corresponding author. Email: scushing@caltech.edu

## Abstract

Transient X-ray absorption techniques can measure ultrafast dynamics of the elemental edges in a material or multiple layer junction, giving them immense potential for deconvoluting concurrent processes. However, the interpretation of the photoexcited changes to an X-ray edge is not as simple as directly probing a transition with optical or infrared wavelengths. The core hole left by the core-level transition distorts the measured absorption and reflection spectra, both hiding and revealing different aspects of a photo-induced process. In this perspective, we describe the implementation and interpretation of transient X-ray experiments. This description includes a guide of how to choose the best wavelength and corresponding X-ray sources when designing an experiment. As an example, we focus on the rising use of extreme ultraviolet (XUV) spectroscopy for understanding performance limiting behaviors in solar energy materials, such as measurements of polaron formation, electron and hole kinetics, and charge transport in each layer of a metal-oxide-semiconductor junction. The ability of measuring photoexcited carriers in each layer of a multilayer junction could prove particularly impactful in the study of molecules, materials, and their combinations that lead to functional devices in photochemistry and photoelectrochemistry.

## Introduction

To understand photochemical and photoelectrochemical processes, a complete measurement of excitation, thermalization, transport, and recombination of charge carriers is needed.[1,2] Dynamics on the femtosecond to nanosecond timescale are often deterministic in defining a device's overall performance.[3–10] Capturing the full variety of coupled electronic and structural dynamics is a challenging task for a single experimental tool. This hurdle is especially true when materials and molecules are combined in the active junctions of solar energy devices.[11–14] Transient X-ray absorption spectroscopy can separate electron and hole dynamics from the vibrational modes that lead to their relaxation and scattering.[15,16] When a sample includes multiple elemental edges, the photoexcited dynamics can be separated by atomic contribution.[15–17] In a multi-element junction, the photoexcited dynamics can be separated into each layer and the transport of charge carriers and vibrational energy can then be mapped throughout a full device.[12–14] Charge selective contacts, plasmonic or hot carrier junctions, light absorber/catalyst interfaces, and their reaction products can be studied in their entirety.

Transient X-ray spectroscopy also brings unique challenges for interpreting experimental data.[18] The X-ray pulse initiates a core-level transition that leaves behind a core hole, which then acts as a perturbation to the final transition state.[18] The formation of this core hole excitons can obscure electron and hole occupations. The core hole exciton can also allow structural information to be inferred without X-ray diffraction experiments. The changes to the photoexcited X-ray absorption are easier to interpret for a molecular sample because the narrower energy valence states are more defined. In a crystal, the complexity of the valence and conduction bands plus screening and many-body effects obfuscates the measurement, placing a higher demand on theory for proper interpretation.[19,20] While multiple theoretical approaches exist for predicting the ground state X-ray structure accurately,[18–21] the prediction of excited state dynamics over femtosecond to nanosecond timescales is still an ongoing development.[22]

In this perspective, we describe the capabilities and challenges of transient X-ray absorption and reflection spectroscopy. We hope to motivate new researchers in relevant fields to consider the technique when studying complex photoexcited dynamics, while also giving an up-to-date picture on how to interpret measured photoexcited changes. In particular, we highlight transient extreme ultraviolet (XUV) spectroscopy because its measurement of delocalized valence dynamics has led to growing popularity for measuring solar energy materials.[4,12–14,23–25] We discuss several representative cases – polarons,[4,10] hot carriers,[23–26] and nanoscale junctions[12–14] – where ultrafast electronic and structural dynamics determine macroscopic device function on much longer timescales. Our goal is to help the reader decide which transient X-ray absorption experiments could be most helpful, while also providing fundamental insight into why photoexcited X-ray absorption edges appear the way they do.

## Background

Photoexcitation raises the energy of an electron, putting it into a non-equilibrium state. The non-equilibrium charge density leads to many-body electronic interactions and couples with vibrational states. The resulting coupled electronic and structural dynamics are the heart of photochemical processes.[1,2] Ultrafast charge carrier dynamics do not necessarily affect the overall efficiency of a solar energy device. Often, carriers thermalize and couple to vibronic states on sufficiently short timescales that their effects on the overall device performance can be ignored. Other times, as is the case for polaron formation or inter-system crossings, the coupled electronic-structural dynamics on ultrafast timescales impact overall device functionality.[3–5,10,27] Most next-generation solar energy devices attempt to extract more energy per incident photon by relying on the ultrafast non-equilibrium dynamics instead of ignoring them.[3,6,7] For example, non-equilibrium vibronic states can tune the selectivity of certain reactions and products in molecular catalysts.[28,29] The excess kinetic energy of photoexcited hot electrons and holes can boost device efficiency.[6,7] The light-induced hybridization or creation of new states could potentially drive hydrogen evolution or $CO_2$ reduction.

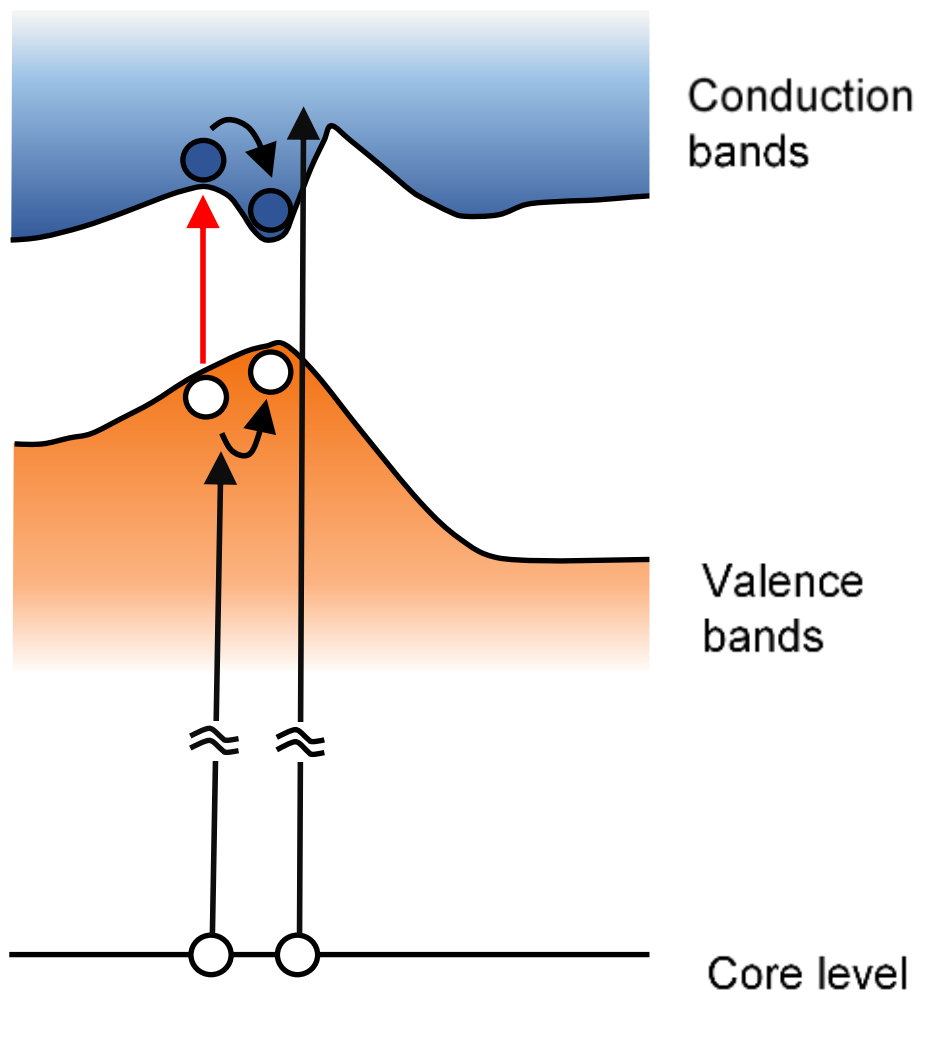

**Fig. 1. Overview of transient X-ray spectroscopy.** Simplified depiction of a time-resolved X-ray absorption experiment. The red arrow indicates transitions with optical pulses, while the broken black arrows indicate an X-ray transition from a core level. The optical pulse excites valence electrons and then the XUV pulse probes the ensuing photodynamics. X-ray pulses can independently probe the electron and hole dynamics.

Ultrafast spectroscopy describes a class of pulsed laser-based techniques that are used to measure photoinduced ultrafast dynamics.[30] Generally, one or multiple laser pulses photoexcite a sample, followed by another laser pulse that probes the changes in the absorption or reflection of the sample. The probe pulse measures the electronic and vibrational changes from the photoexcitation. Varying the time delay between the pump and probe pulses enables the real-time observation of photoexcited dynamics. The ideal measurement approach would be to pick a probe wavelength that matches the transition state or resonant feature of interest. However, when using ultraviolet, visible, or infrared light as a probe, overlapping spectral features in the valence states can make it difficult to deconvolute the electronic and vibrational dynamics, especially in multi-element compounds or integrated multi-layer samples, which are common in photoelectrochemistry.[11] Electron and hole dynamics are also difficult to separate in the signal because their joint density of states is measured. Further, it is helpful to isolate the photoexcited dynamics in terms of different dopants or atomic species. Transient X-ray spectroscopy aims to overcome these issues by probing a core-to-valence transition instead of the valence-valence transition or a vibrational resonance itself (Fig. 1).

The most important step in designing a transient X-ray experiment is selecting the X-ray probe energy. Core-level transitions span energies from tens of eV to tens of keV. Every element has multiple edges (or core-to-valence transitions) within this broad energy range. Probing different transitions of the same element measures different information about the photoexcitation. High energy hard X-rays at several keV (sub-nm) give atomic-site specific electronic dynamics as well as direct structural information.[31] Comparatively, low energy XUV photons around 100 eV and below (10-100 nm) probe the delocalized valence dynamics (including electron or hole energies) but can only infer structural information.[4,13,24–26] Soft X-rays, which describe the hundreds of eV to keV range, are a compromise between the information in XUV and hard X-ray measurements.

There are various experimental methods for generating ultrashort X-ray pulses for transient measurements. An X-ray free-electron laser (XFEL) or a synchrotron facility can generate hard X-rays.[32] XFEL, synchrotron, and tabletop XUV experiments start to overlap at intermediate soft X-ray energies from 300 eV to 1 keV.[33] Ultrashort, coherent XUV pulses are generally created from a table-top setup using a scheme known as high-harmonic generation (HHG).[34] Whereas synchrotrons or XFELs modulate relativistic free electrons to create X-rays, HHG uses the strong electric field of an optical pulse to photoionize, accelerate, and recombine electrons from a noble gas atom to create X-rays.[34] Since HHG is a coherent process, the spatial and temporal characteristics of the driving pulse are preserved in the XUV pulses that are generated.[35,36] The resulting broadband XUV pulse duration can be significantly shorter (attoseconds) than the driving optical pulse (femtoseconds).[35] The broad bandwidth of HHG emission makes it useful for measuring multiple atomic edges in one measurement.

A few comments should be made on the practical challenges of table-top X-ray sources. At XFELs and synchrotron sources, experimental difficulties center around limited access time and the inability to construct extensive experiment-specific apparatus at the facility. For table-top XUV experiments, on the other hand, unlimited access is traded for low X-ray fluxes. In general, the HHG process has a conversion efficiency of $10^{-5}$~$10^{-7}$ with commonly used noble gases and <20 fs optical driving pulses. Even with high intensity lasers, the produced XUV flux from HHG is limited to the pJ to nJ range.[37] The low conversion efficiency of the HHG process and its reliance on the peak intensity adds to the cost and complexity of the experiment. The low flux and short absorption depth of XUV radiation (~100 nm for p-block and ~10 nm for d-block elements) also necessitate special sample preparation. These samples can include transmission measurements on a silicon nitride or diamond window,[4,24,25] reflection measurements on smooth solid samples,[10,13,38] or sub-micron liquid jet methods.[39] Signal to noise ratios of a few mOD can be achieved with multiple hours of data acquisition using an XUV CCD spectrometer. Better signal to noise ratios are possible through lock-in detection with an XUV photodiode,[40] but the better signal to noise ratio must be balanced against the temporal demands of having to take multiple scans at different energies. Perturbations to X-ray transitions span the complete edge.[4,20] This photoinduced renormalization of X-ray spectra over a broad energy range in response to small perturbations makes XUV spectroscopy a sensitive tool for detecting small changes to local chemical environment.

## Application in solar energy materials

### Photocatalytic reactions limited by local charge-trapping (small polaron formation)

Small polaron formation occurs when an electron or hole interacts with a polar lattice to localize the respective wavefunction at a single atomic site.[4,5,10] The localization occurs through a distortion in the surrounding bonds. The transformation from a free electron-like wave to a trapped, defect-like charge severely limits charge carrier mobility and increases recombination events. There is a growing belief, backed by transient X-ray measurements, that photoexcited small polaron formation explains why metal oxide photocatalysts have never reached the photoconversion efficiencies suggested by their band gap.[4,10] Small polarons are especially prevalent in transition metals with partially-filled d-orbitals because of their localized nature (Fe containing compounds, for example). The localized d-orbitals allow efficient molecular catalysis but also have a small energy cost for the formation of a small polaron. For largely unfilled or nearly fully filled d-orbitals (e.g. $TiO_2$ and ZnO), large polarons are formed.[41] Large polarons refer to the fact that the trapped charge and the associated lattice distortion is extended over multiple unit cells to the point where it is only a slight impediment to charge transport.[41–44] Photoexcited polaron formation is still an active area of research, and many intermediate cases such as $BiVO_4$ remain debated.[45] It is becoming increasingly apparent that the photocatalysis community must avoid or reduce small polaron effects to create efficient photocatalysts.

As an example, recent transient XUV absorption measurements of the Fe $M_{2,3}$ edge of hematite confirm the formation of small polarons around Fe sites after optical excitation.[4] Fig. 2A illustrates the small polaron formation in photoexcited α-$Fe_2O_3$. Optical excitation induces electron transfer from a negatively charged oxygen site (dark orange) to the positively charged Fe center (light orange). The charge transfer process reduces the Coulomb interaction between the Fe center and the surrounding O atoms. The reduced Coulomb interaction causes a local lattice expansion to give electron localization at the Fe center. The XUV core-level transition of the Fe site is sensitive to its surrounding environment through the core hole perturbation of the transition state. As will be discussed later, the Fe $M_{2,3}$ edge is dominated by angular momentum coupling of the core hole to the valence states. The coupling results in multiple peaks that are re-arranged from the unoccupied valence orbitals (Fig. 2B). The charge transfer excitation better screens the core hole perturbation, so the energy separation between said peaks reduces, producing the blue spectrum in Fig. 2B immediately after photoexcitation. The polaron formation creates a local anisotropic lattice distortion that breaks the degeneracy of Fe 3p core level. The sub-picosecond shift in the differential absorption (Fig. 2B, bottom panel) corresponds to the resulting change to the XUV absorption (Fig. 2C). Hence, the polaron formation rate relative to the charge transfer state is apparent in the differential absorption, as shown in Fig. 2C and D.

After decomposition of the differential absorption, the small polaron formation is measured to occur in under 100 fs (Fig. 2D), matching the first electron-phonon scattering events. Importantly, the small polaron state persists until the photoexcited electron and hole recombine. The few picosecond polaron formation will therefore determine any later photocatalytic steps. Interestingly, polaron formation has been measured not to be influenced by grain boundaries or defect states. Small polaron formation is an intrinsic property of the lattice that cannot be removed by better quality materials alone. Bulk and surface polaron formation occur on slightly different timescales, with bulk polaron formation occurring with a time constant of 90 fs and surface polarons forming with a time constant of 250 fs.[4,46] Molecular functionalization of the hematite surface has also been shown to change the lattice reorganization and polaron stabilization energies, but polarons still form with rates between 146 fs for OMe-PPA-$Fe_2O_3$ and 250 fs for hydroxyl-$Fe_2O_3$.[46] Furthermore, comparing small polaron formation in hematite and goethite (a-FeOOH) has shown that small changes in ligands and crystal structure have a significant effect on formation rates, with polarons forming in 180 ± 30 fs in goethite nanorods and 90 ± 5 fs in a hematite thin film.[47] Engineering materials to remove polaronic effects remains outstanding questions for photocatalysis and photoelectrochemistry.

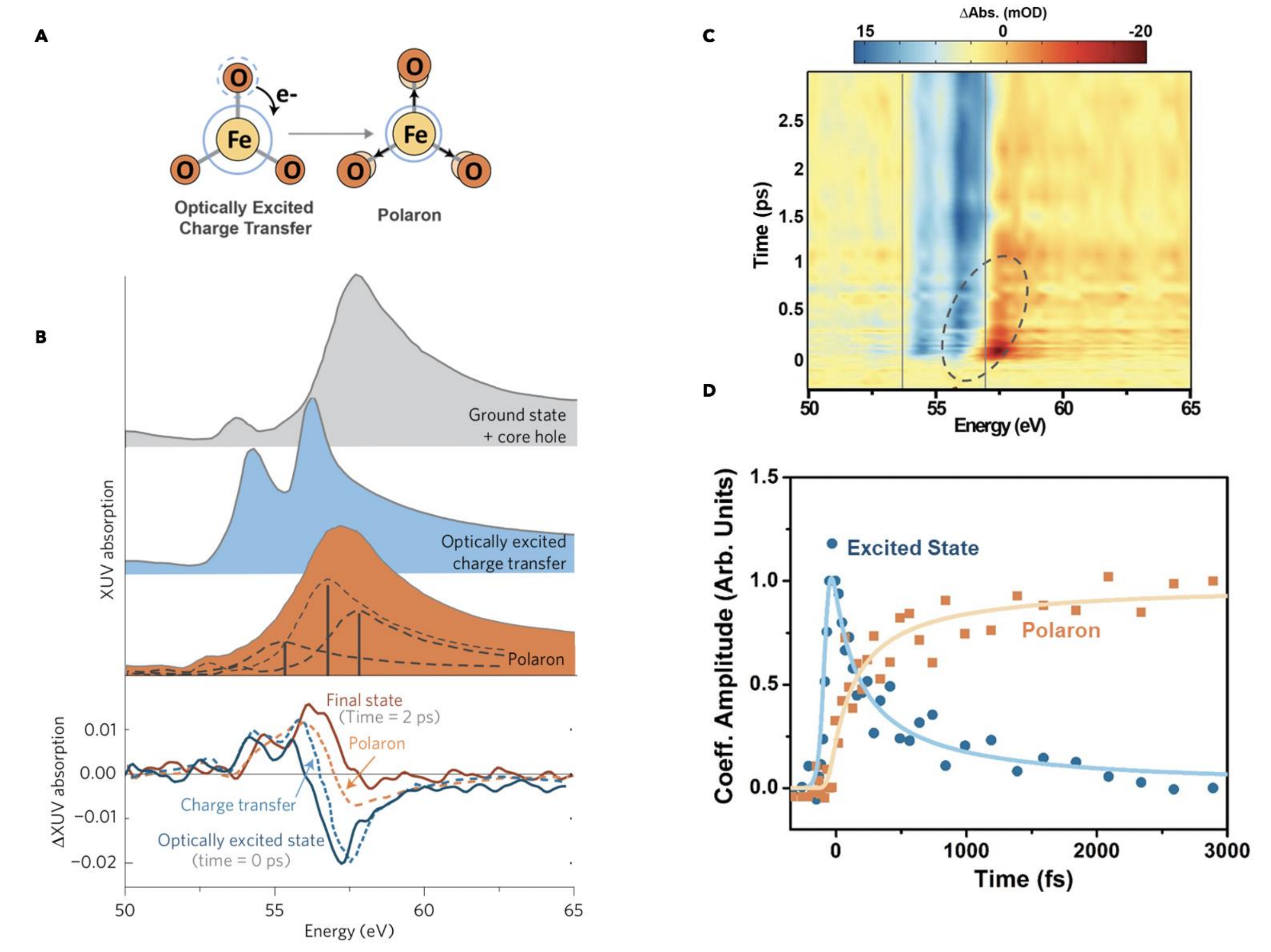

**Fig. 2. Use of XUV to probe small polaron formation.** (A) Initial excitation of $Fe_2O_3$ leads to a charge-transfer transition from oxygen (dark orange) to iron (light orange). The initial excited carriers undergo optical-phonon scattering, during which the charge can be trapped by forming a small polaron and expanding the lattice locally. (B) XUV absorption spectra of the polaron formation process. The top panel is experimental data; the middle two panels are simulations. Dashed lines represent the contributions of the multiplet splitting and core hole absorption to the predicted small polaron spectrum. The final panel shows the predicted contributions of charge transfer and the polaron to the differential absorption spectrum (dashed lines) compared to the time = 0 ps and time = 2 ps differential absorption measurements following a 480 nm excitation pulse (solid lines). (C) Transient absorption measurement. Following excitation, the differential absorption reveals regions of increased and decreased absorption around 55 and 56 eV. The energy of the zero crossing between the areas of increased (blue) and decreased (red) absorption shifts on a ~500 fs timescale and is circled for clarity. (D) The experimental time = 0 ps and time = 2 ps slices are used to extract the amplitude of the charge-transfer and small polaron state using a multivariate regression of the experimental data. These amplitudes are fitted with a rate equation model for small polaron formation (solid lines). Fig. is adapted from Reference 4.

## Energy-resolved electron and hole kinetics

As highlighted by small polaron formation, controlling the relaxation pathways of photoexcited carriers is crucial for better optimizing solar energy devices. Efficient extraction of excess kinetic energy of hot (non-thermalized) electrons and holes can overcome the theoretical efficiency limit.[3,6,7] Hot electrons and holes created by plasmonic nanoparticles can create new photoproducts not reachable with thermalized carriers.[8] The ability of transient X-ray absorption spectroscopy to separately measure the electron and hole energies as a function of time is very useful for these applications. Measuring electron and hole transport across interfaces is critical for carrier selective contacts, whether using thermalized or non-thermalized photoexcited carriers. Transient X-ray absorption measures the momentum summed electron and hole energies unlike the individual bands of angle resolved photoemission spectroscopy (ARPES). However, X-rays do offer advantages over ARPES in terms of penetration depth,[12] measuring electronic and structural dynamics simultaneously,[4,26] element specificity,[17] and the ability to measure insulating or oxidized samples.[48]

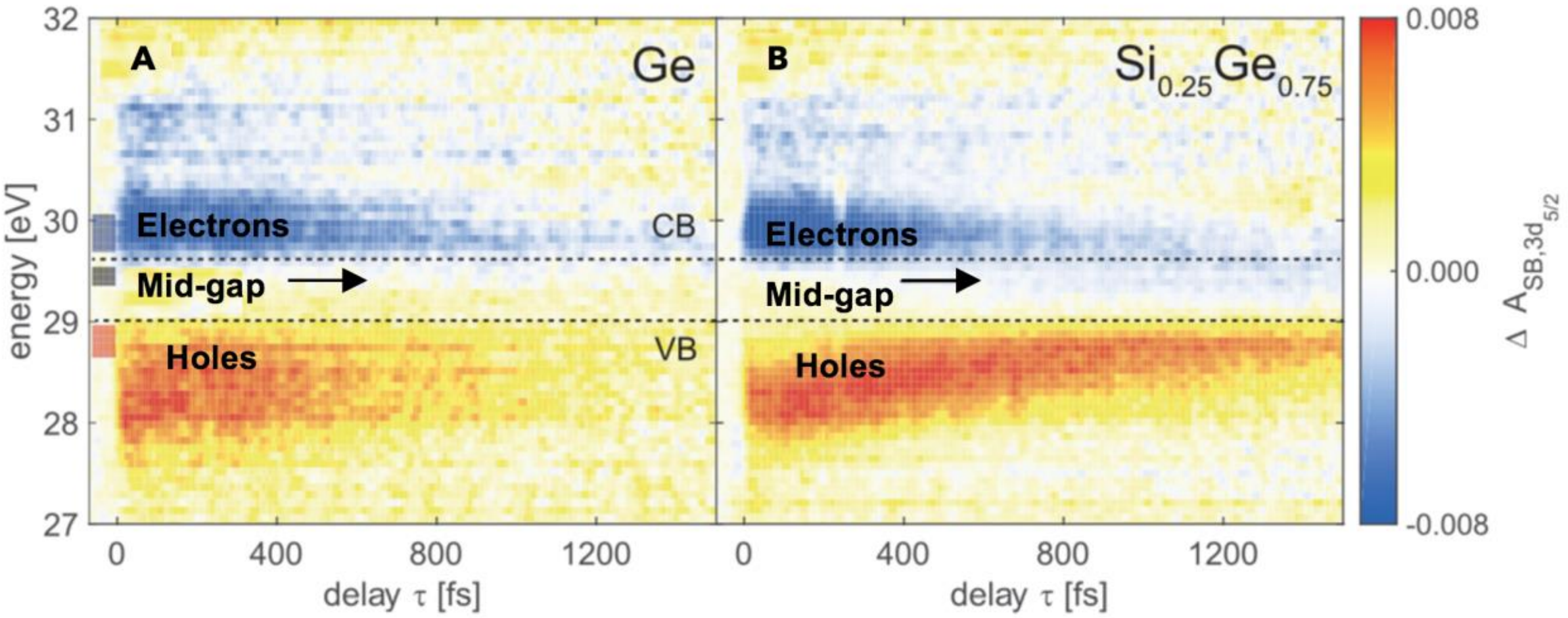

**Fig. 3. Time-resolved state blocking in a germanium nanocrystal (A) and silicon-germanium alloy (B), measured in the XUV.** (A) Carrier dynamics in nanocrystalline germanium reveal the relaxations and decays of photoexcited carriers in the valence band (VB) and conduction band (CB). (B) Direct comparison with silicon-germanium alloy shows that the Ge $M_{4,5}$ edge can be employed to study carrier dynamics in the alloy. In the silicon-germanium alloy, an additional weak feature in the mid-gap (indicated by the black arrow), which is assigned to trap states, can be observed at longer time delays. Fig. is adapted from Reference 25.

Fig. 3 shows the measured transient XUV absorption at the germanium $M_{4,5}$ edge of a germanium thin film[24] (A) and a $Si_{0.25}Ge_{0.75}$ alloy[25] (B). In this experiment, a visible-to-near-IR femtosecond pulse excites the thin film samples. The photoexcitation creates free electrons and holes in the conduction bands and valence bands, respectively. Holes in the valence bands results in new X-ray transitions and show up as a positive (red) contribution. Similarly, photoexcited electrons in the conduction band will block the XUV transitions, reducing the XUV absorption (blue). As a result, electrons and holes dynamics are separated in the transient XUV spectra (Fig. 3A and B). In this figure, a constant energy shift has been removed from the differential absorption to give a better picture of the electron and hole dynamics. A shift of the entire edge to lower energy is measured because the photoexcited state changes the screening of the core hole perturbation (similar to band gap renormalization in optical excitation).

From the processed differential absorption, electrons are measured to thermalize quicker than holes over a one picosecond timescale. The slower hole thermalization is expected because of the higher effective mass of holes. The holes also have a higher initial energy because of where the optical excitation took place within the band structure of Ge and the Si-Ge alloy. The kinetics of the Si-Ge alloy can be compared to measurements of Ge alone (Fig. 3A). After the carriers are thermalized, the filling of the mid-gap defect states in the alloy by the electrons can be measured (Fig. 3B). At this point, the decreasing amplitude in the differential absorption shows that both the electron and hole density start to decrease due to carrier recombination. In addition to carrier recombination, electrons can be trapped in defect states near the conduction band edge, leading to a rise in the signal attributed to the mid-gap state as shown in Fig. 3b. While the defect state density in the Si-Ge alloy is high, improvements in instrument sensitivity could make measurements of few percent dopant and defect states possible. Measuring how a few percent dopant or defect states affect the photoexcited dynamics of an alloy is a promising direction for the transient X-ray techniques.

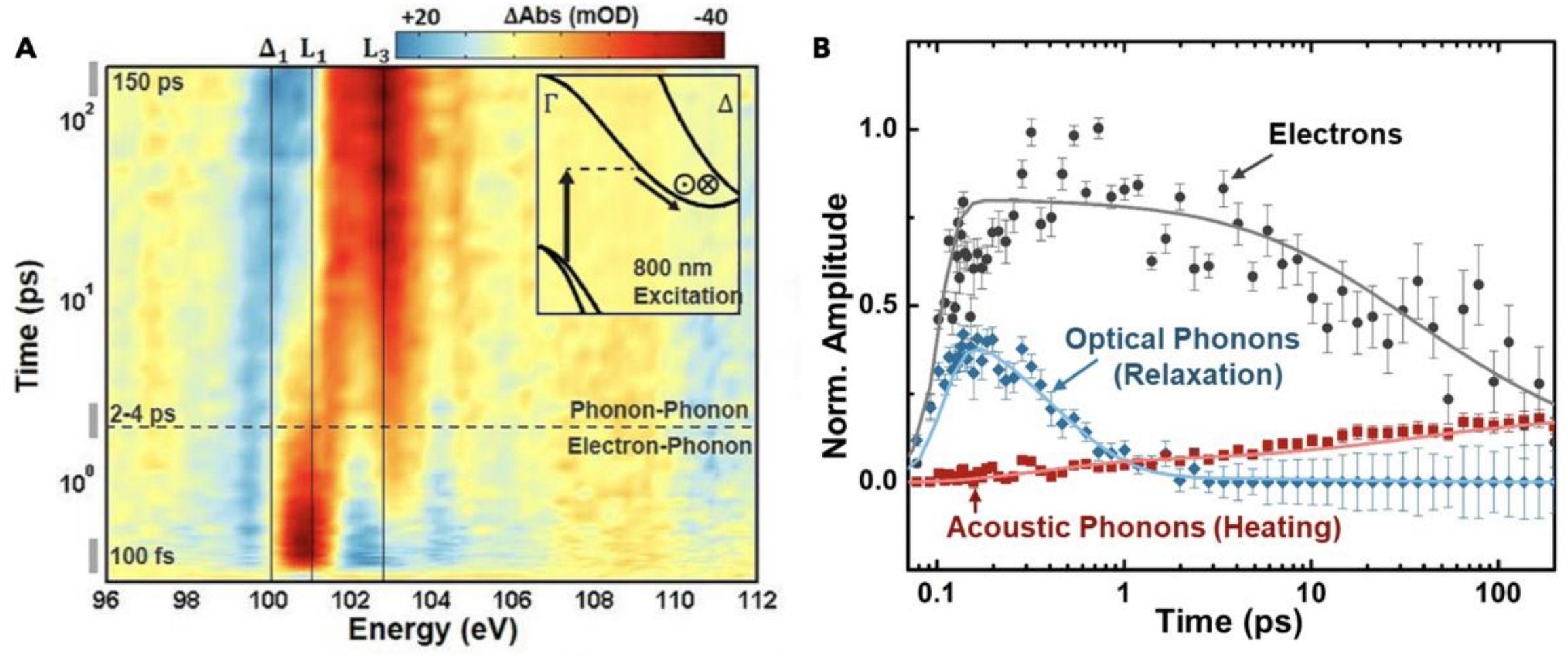

**Fig. 4. Time-resolved XUV spectroscopy of the silicon $L_{2,3}$ edge.** (A) The differential absorption is shown on a logarithmic time scale from 0 to 200 ps for 800 nm excitation. The inset represents some of the possible excitation and scattering pathways for the excited electrons. The in and out of plane arrows indicate where inter-valley scattering between degenerate valleys is possible. The cross-over time between predominantly electron-phonon scattering or phonon- phonon scattering is also indicated by the dashed horizontal line. (B) Time dependent carrier density, lattice deformation, and thermal lattice expansion as extracted from a non-linear fit of the measured excited state XUV absorption in A. The solid lines show the predictions of a three-temperature model (3TM) for the carrier density, optical phonons, and acoustic phonons. The points refer to the quantities extracted from the experimental data, with error bars representing the standard error of the fit to the experimental data. The non-linear fit is weighted by the experimental uncertainty. The scatter of the points is representative of the experimental noise. The lattice deformation is approximated as an equilibrium [100] expansion and the lattice heating is approximated as an isotropic lattice expansion. The log scale of time is offset by 100 fs for visualization. Fig. was adapted from Reference 26.

Carrier kinetics have been measured in a range of materials using transient XUV spectroscopy.[23,49] Fig. 4A shows the differential absorption of Si from which similar information can be obtained. Comparing Fig. 4 to Fig. 3 highlights the trade-off between different core-level transitions. Interpretation of the Si $L_{2,3}$ edge is more difficult because of the higher X-ray energy and thus stronger core hole perturbation effects. However, structural effects are also now more present[26], whereas they are more difficult to interpret in the Ge $M_{4,5}$ edge at lower energy. For the more complex Si experiment, the electron dynamics can still be extracted via theoretical modeling of the transition state. Here, a Bethe-Salpeter equation (BSE) approach[20] was used to deconvolute the transient XUV spectrum into both structural and electronic changes. The electron population versus time, the thermalization of the electrons, and the acoustic phonon generation can then be extracted from the differential absorption spectrum (Fig. 4B).[26] The extracted kinetics match those of known silicon scattering rates from optical measurements, but now the electron and phonon dynamics are measured separately. Valley-dependent scattering rates can also be obtained by changing the excitation wavelength in the measurement. The measurements of the Si $L_{2,3}$ edge are good examples of where intuitive interpretation based on increased and decreased absorption fail without modeling. Once modeled, however, future experiments can be more easily analyzed with simpler decomposition or lineout methods.

### Charge transfer in multilayer devices

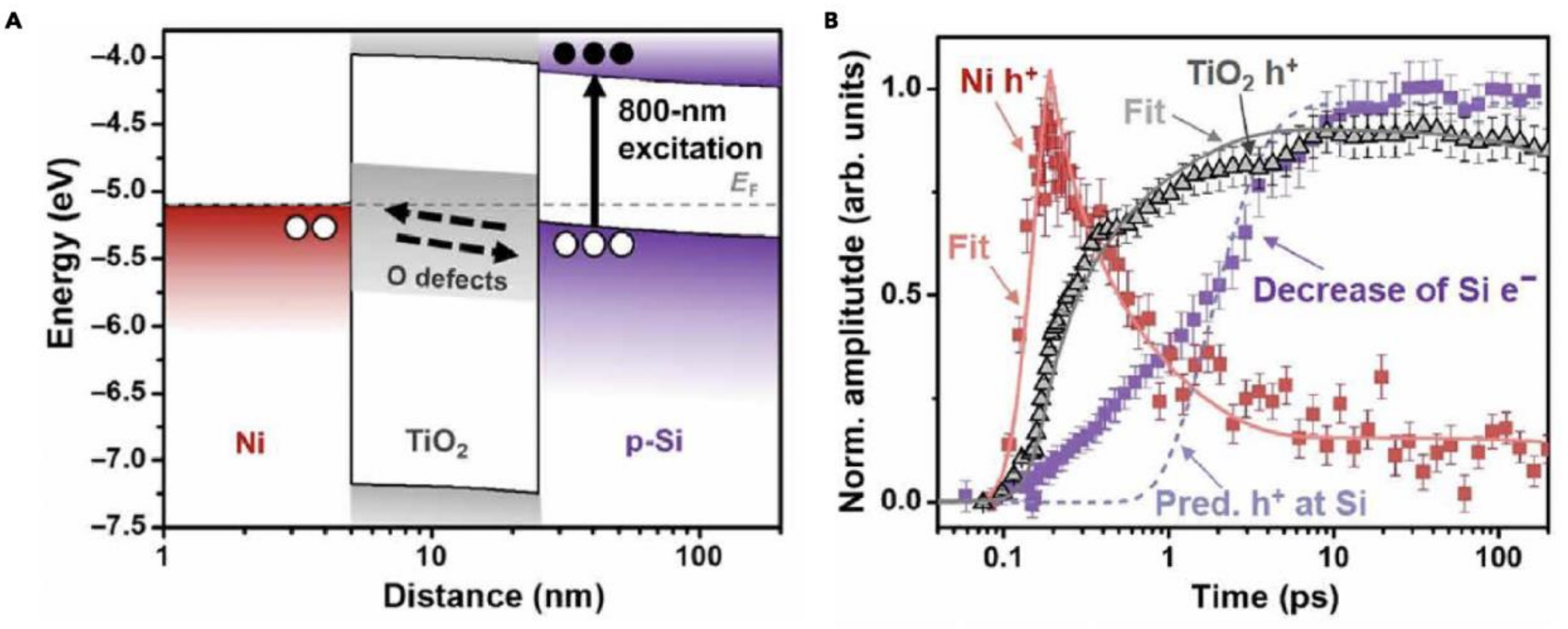

**Fig. 5. Characterization and measurement of the Ni-$TiO_2$-Si junction.** (A) The energy level alignment for the metal, oxide, and semiconductor is shown, along with the expected photoexcited hole transfer in the p-type metal-oxide-semiconductor junction. (B) The edge shift kinetics, which indicate the hole kinetics, measured for the $TiO_2$ (gray triangles) and the Ni (red squares) in the junction are compared to the increase in electron recombination (i.e. decrease in electron signature at the Si edge) in the junction (purple squares). The solid line is a fit to the diffusion equation with a diffusion constant of 1.2 ± 0.1 $cm^2/s$ and a surface recombination velocity of 20 ± 50 cm/s. The measured decrease in electron density qualitatively tracks the diffusion of holes through the $TiO_2$. The dashed line represents the predicted (Pred.) arrival of holes at the Si-$TiO_2$ interface based on the fit diffusion kinetics. Fig. was adapted from Reference 12.

Solar energy devices rely on multi-materials junctions to separate and transport charges.[50] The critical interfaces include charge-selective contacts in a photovoltaic cell or the semiconductor-catalyst-electrolyte interface of a photoelectrochemical cell.[11,50] At the microscopic level, multiple charge transport events, oxidation-reduction reactions, and mass transport are balanced at the interface from femtosecond to second timescales.[1,8,11,50] Measuring the ultrafast photodynamics of buried interfaces is therefore a challenging, but valuable spectroscopic task. While charge transfer is readily measured in two-component systems[17,51] (such as between solids and quantum dots or molecules), adding the metal contacts and multiple broadband absorbers of a full solar energy device makes the measurement non-trivial. Element-specific transient X-ray measurement can uniquely track the photoexcitation and electronic-structural coupling that leads to both charge and heat transport.[23,26] The X-ray penetration depth can be varied as a function of incident angle for probing different layers of junctions or to differentiate surface and bulk dynamics. The measured kinetics can then be transformed into transport properties if the sample geometry is known from TEM or SEM images. The ultrafast measurements are then relatable to the results of device averaged bulk electrical measurements.[12]

As an example, Fig. 5A shows the carrier transport processes in a p-Si/$TiO_2$/Ni multilayer junction.[12] In this experiment, a near-infrared pulse was used to photoexcite the Si. Due to the band alignment in the junction, photogenerated holes in the Si layer then transport ballistically to the Ni layer. Spectroscopically, this transport is observed as an energy shift in the Ni $M_{2,3}$ edge energy in the first ~ 100 fs. Picoseconds later, holes in the Ni diffuse back through the $TiO_2$ to the Si. This back-diffusion of holes increases the Ti oxidation state, which is observed by an energy shift in the Ti $M_{2,3}$ edge energy. Eventually the holes recombine with electrons, the loss of which is measurable by analysis of the Si edge with the BSE approach as described in the previous section. Carrier transport kinetics in this multi-junction device are therefore fully characterized (Fig. 5B). The holes arrive on the Ni edge after a 33 ± 8 fs delay from the photoexcitation pulse. The $TiO_2$ layer is 19 nm thick, so the holes have a tunneling velocity of $(5.8 \pm 1.4)\times10^7$ cm/s. This speed is similar to the velocity of the electrons in the Si, indicating ballistic tunneling. Given the built-in field of the junction, the hole mobility can then be estimated as 390 ± 100 $cm^2$/V, which is comparable to the established value of $10^{15}$ per $cm^2$ p-doped Si (450~500 $cm^2$/V·s).[52] A charge transfer efficiency of 42% after initial photoexcitation was also extracted from measurement. The surface recombination velocity at the Si-$TiO_2$ interface is measured to be >200 cm/s.

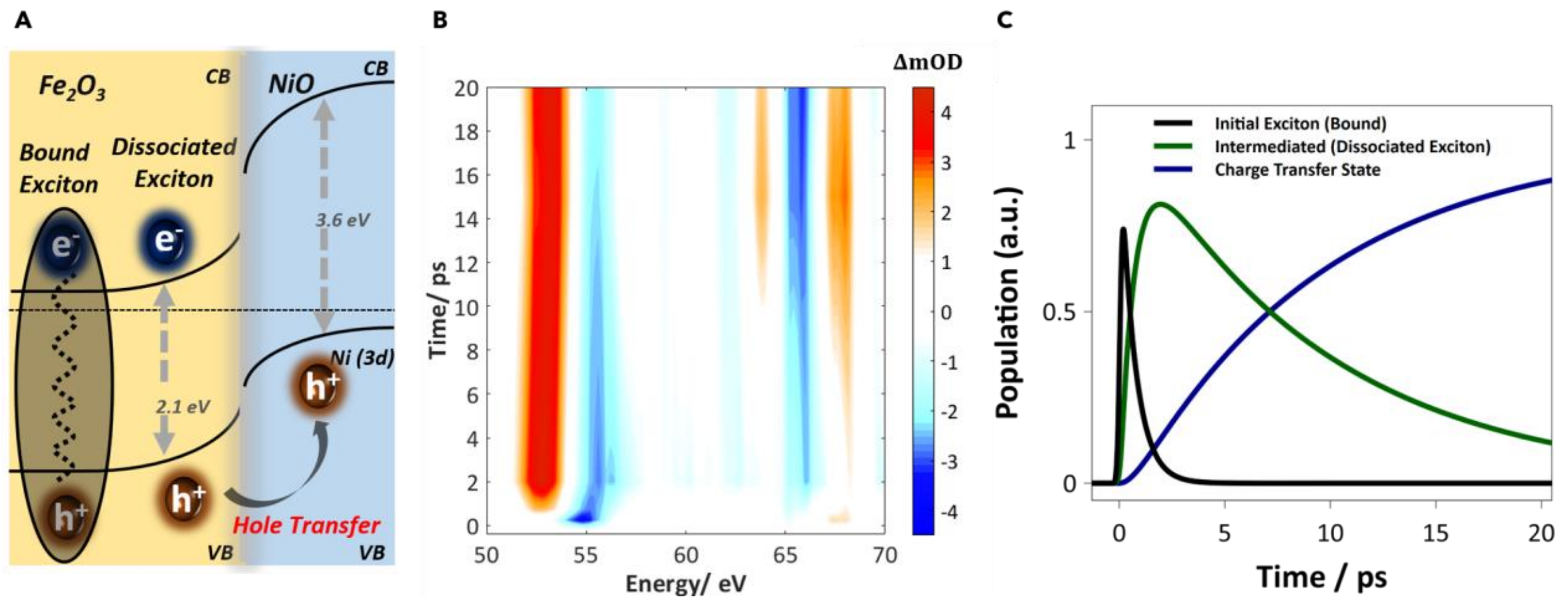

**Fig. 6. Direct observation of exciton dissociation and interfacial hole transfer across a $Fe_2O_3$/NiO heterojunction with XUV reflection-absorption spectroscopy.** (A) Schematic depiction of a two-step sequential mechanism of hole transfer. The interfacial field drives the exciton dissociation in the space charge region of $Fe_2O_3$/NiO heterojunction. The subsequent hole transfer occurs via the drift of the carrier across the interface. (B) Contour plot showing the experimental transient reflection–absorption data of $Fe_2O_3$/NiO heterojunction excited with 400 nm light. (C) The population of the bound exciton, dissociated exciton, and hole transfer states is obtained from the global fit of data in B. Photoexcitation at 400 nm produces a charge transfer exciton in $Fe_2O_3$. This charge transfer exciton dissociates to form dissociated exciton with a rate constant of $k_1$ (680 $\pm$ 60 fs). The subsequent interfacial hole transfer across the heterojunction to NiO valence band occurs with a rate constant of $k_2$ (9.2 $\pm$ 2.9 ps). Fig. was adapted from Reference 13.

Similar interfacial hole transfer dynamics have been reported in other solar junctions, including $Fe_2O_3$/NiO and Zn/n-GaP junctions.[13,14] The differential absorption measured in the $Fe_2O_3$/NiO heterojunction using transient XUV reflection-absorption spectroscopy (Fig. 6) highlights the formation of a transient $Ni^{3+}$ state that forms ~ 10 ps after photoexcitation, indicating the importance of the 3d character of the NiO valence band states.[13] Furthermore, the simultaneous measurement of the dynamics at both the Fe $M_{2,3}$ edge and the Ni $M_{2,3}$ edge (Fig. 6B) shows the two-step nature of the mechanism. An initial exciton dissociation occurs in $Fe_2O_3$ 680 fs after photoexcitation, and subsequently the hole is injection into NiO after approximately 9.2 ps. Fig. 6C illustrates the kinetics of this process, starting with a bound exciton, followed by the dissociated exciton until finally the hole transfer state dominates. Overall, transient XUV spectroscopy gives a detailed microscopic picture of charge transport in multi-layer junctions.

**Interpreting photoexcited X-ray absorption spectra**

The interpretation of photoexcited X-ray absorption spectra requires theoretical modelling to accurately extract charge carrier and structural dynamics, regardless of the energy range used.[18–21] An X-ray probe photon excites a core electron to a valence state, leaving a positively charged localized hole in the core state. As a first guess, the transition probability of the X-ray, and thus the measured absorption spectrum, would be proportional to the unoccupied valence density of states given the narrow energy of the core level. However, the core hole that is created during the X-ray transition perturbs the final transition valence density of states.[18,20,21] The strength and extent of this perturbation determines what information can be extracted from the transient X-ray absorption experiment. In other words, the core hole perturbation is the defining link between the measured change in the X-ray peaks and the photoexcited information that can be extracted from the measurement.

The general ranking of core hole perturbations, from strongest to weakest, is angular momentum effects (atomic multiplet splitting and spin-orbit coupling), ligand field effects on these angular momentum perturbations, and the creation of core hole excitons from the X-ray excited electron. The ability of the valence orbitals to screen the core hole perturbation determines the exact peak amplitudes and spectral shifts within this general ordering.

For the elements with localized d-orbitals, angular momentum and spin-orbit effects are dominant in X-ray spectra. Atomic multiplet coupling between the core level and valence states leads to a complete redistribution of the unoccupied valence structure across tens of eV. The angular momentum effects also make extracting exact electron and hole energies more difficult. Screening and ligand field effects determine the exact strength and separation of the peaks. A lower energy peak cannot reliably be assigned to a lower energy valence orbital, such as the $t_{2g}$ or $e_g$ density of states (Fig. 7A). Instead, theoretical methods[19–21] are used to map out to which states the X-ray peaks correspond. A change in oxidation state is usually obvious, such as the ligand to metal charge transfer[4] in Fig. 2, but the electrons and holes are not as spectrally resolved as in Ge (Fig. 3A). However, the strong valence to core hole coupling does provide some advantages. Slight changes in electronic or coordination environments after photoexcitation become exaggerated in the X-ray spectrum.[4] Changes in oxidation state and structural dynamics such as polarons are readily apparent.[4,10,46] The same can be said for f-block elements, but now to an even greater extent because of their increased spatial localization. As a result, f-orbital X-ray transition spectra are often dominated by separated, narrow linewidth peaks as compared to a single lumped peak.[53]

For elements with s- and p-orbital valence electrons, the angular momentum coupling is moderate. The primary perturbation affecting the X-ray absorption structure is the creation of core hole excitons. The core hole exciton is usually narrow in linewidth and shifted lower in energy from the expected orbital. In a molecule, the core hole exciton is quite pronounced. In a solid, the -delocalized s and p valence orbitals can screen the core hole to weaken the excitons amplitude. The X-ray spectra then more closely (but not exactly) match the density of states in the material (Fig. 7B).[23] Changes in occupation of electrons and holes in specific bands can therefore be directly measured or extracted.[24,25] Electrons block transitions while holes allow new transitions, leading to negative and positive peaks in a transient X-ray spectrum[24,25] (Fig. 3A). The change in valence occupation thus changes the screening of any core hole excitons. The resulting renormalization can cause spectral changes across the whole X-ray absorption, as seen in the Si $L_{2,3}$ edge in Fig. 7C.[23,26]

However, intuition about screening of the core hole does not always hold true. An important case is the X-ray absorption spectrum of metals.[54] For a metal, many-body interactions between the core hole and free electrons dominate the X-ray absorption. The renormalization from the core hole exciton in the partially-filled d-valence bands creates an exponentially increasing density of states near the Fermi level. The X-ray absorption spectrum is therefore broadened and redistributed from the underlying empty orbitals, even though atomic multiplet effects are minor (Fig. 7C).

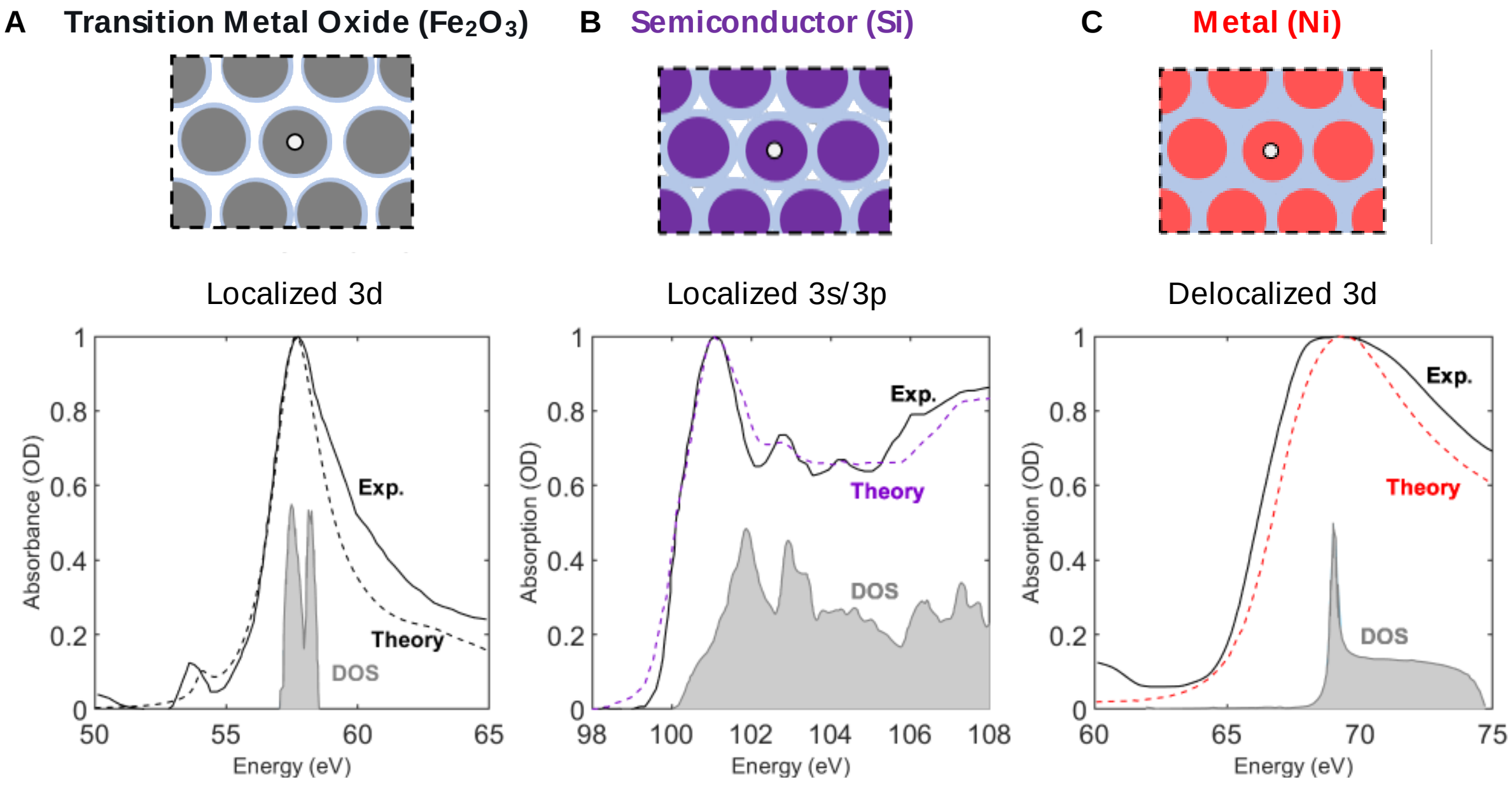

**Fig. 7. The XUV absorption in solid state materials follows the available density of states to varying degrees depending on the screening of the core hole perturbation.** (A) In a metal oxide, the localized d levels interact strongly with the core hole of the XUV transition. The final transition density of states is dominated by atomic multiple splitting and is therefore distorted from the unoccupied valence states. (B) In a covalent semiconductor, the delocalized electrons shield the core hole perturbation, and the measured experimental absorption is close to the unoccupied density of states. (C) In metals, the white edge effect results in vanishing ground and excited state overlap at the Fermi level, resulting in large differential absorption relative to the available density of states. Fig. was adapted from Reference 12.

In all cases, as the photoexcited carriers relax by coupling with vibrational modes and phonons, heating features show up in the transient X-ray spectra.[26] Heating features refer to the lattice expansion that exists from the anharmonicity of the excited vibrational modes. The expansion of the lattice decreases the bond hybridization and band gap of a material because the orbital overlap is decreased. Similarly, the interaction of the core hole with surrounding atoms is decreased, and the overall X-ray spectrum is generally shifted to lower energy. This behavior is prominent on longer time scales (tens to hundreds of picoseconds) than the initial carrier effects and is present for almost any X-ray energy range.[26] The usefulness of the heating feature, beside studying heat transport on nanometer scales, is that its kinetics can be compared to the carrier relaxation kinetics to infer carrier-phonon and phonon-phonon scattering timescales. Phase changes and other lattice distortions are similarly measurable through this manner.

The Bethe-Salpeter Equation (BSE) and time-dependent density functional theory (TDDFT) calculations can accurately account for the effects of the core hole perturbation.[20] First, TDDFT calculations can be used for simulating molecules and materials on the sub-picosecond timescale.[55] This approach gives the real time evolution of the state, predicting attosecond dynamics, but is computationally expensive for longer timescales. In the second approach, the BSE equation models experimental X-ray spectra from interactions between valence electrons and the core hole (Fig. 8A).[20] Here, a DFT spectrum is used to predict the ground state valence orbitals and filling. An optional GW step is used to correct the DFT spectrum for electron-electron interactions. Next, the core hole excitons and any angular momentum coupling is calculated using BSE. Fig. 8B shows a bubble plot of the amplitude and k-space distributions of the core hole-excitons calculated for the Si $L_{23}$ edge. Time series of the excited state spectra are then calculated on the assumption that the core hole excitation (attoseconds) is slower than carrier or phonon dynamics (femtoseconds and longer). The predictions of the transient X-ray spectrum are then calculated by the BSE incorporating the changes from a density functional perturbation theory (DFPT) method with the Boltzmann transport equation.[22] The effect of electron and hole distributions, as well as phonon occupations, can therefore be predicted or extracted from experimental measurements.

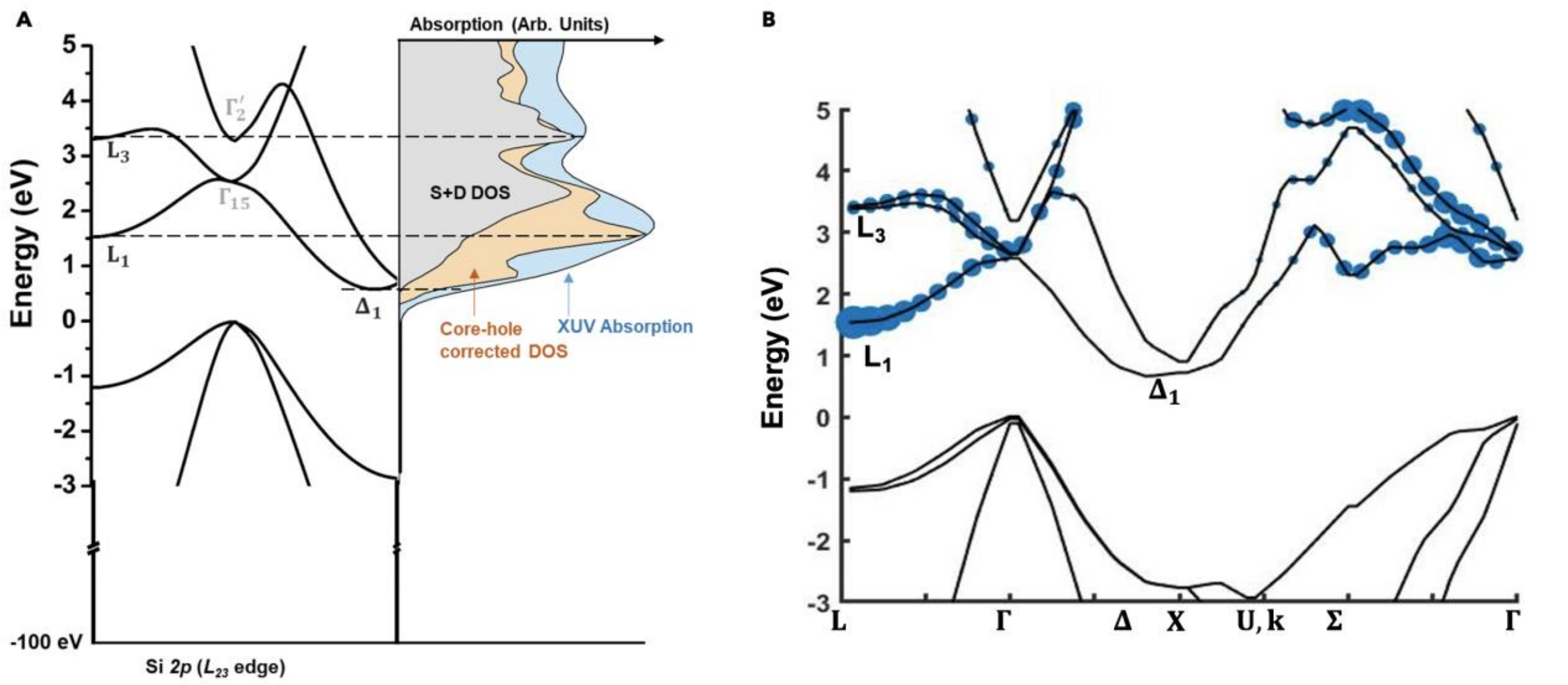

**Fig. 8. Modeling ground state absorption XUV absorption in silicon with the Bethe-Salpeter equation (BSE) approach.** (A) The BSE-based model provides good agreement with the experimental XUV absorption and theory when correcting the density of states (DOS) transition with the effects of the core hole perturbation. Without this core hole correction, there is significant deviation between the experimental measurement and DOS. (B) By applying the BSE approach across silicon's band structure, the core hole exciton transition density can be calculated to indicate which high-symmetry points have the largest contributions to the XUV spectrum. The BSE approach also allows for different components of the excitonic Hamiltonian to be separated contributions from spin-orbit, multiplet splitting, higher order processes such as bubble/ladder interactions, and long-range effects. Fig. was adapted from Reference 23.

## Outlook

This perspective outlines state-of-the-art applications of transient X-ray spectroscopy, particularly low-energy XUV measurements, for understanding solar energy materials. To date, only a few solar driven dynamics have been measured, and the application of transient X-ray spectroscopy to solar energy materials is still nascent. An important next step is to include *in situ* and environmental cells to match the cutting edge of synchrotron experiments. The ideal measurement would probe photoexcited dynamics of carriers and quasiparticles, charge transfer between the electrode-electrolyte interface, and the associated chemical reaction intermediates and products that form. For example, recent static *in situ* measurements were taken of the electrochemical $CO_2$ reduction reaction on Cu catalyst with X-ray absorption spectroscopy and X-ray diffraction.[56] *In situ* measurements are more difficult with table-top XUV experiments because of the limited penetration depth and lower photon energy range. However, table-top sources are now pushing into the water window (282~533 eV) with sufficient flux for transient experiments in the liquid phase.[33] This energy coverage allows the study of organic elements in addition to the more easily accessible inorganic ranges integrated in photoelectrochemical liquid cells. Transient X-ray spectroscopy is also significant for the insight it can give into alloyed and multiple-element compounds critical to these photoelectrochemical cells. Signal-to-noise ratios are reaching the levels needed to measure the role of defect and dopant state dynamics.

In addition to linear X-ray absorption spectroscopy, novel nonlinear X-ray methods are emerging as a promising field for both fundamental light-matter interaction and material characterization.[57] Recent developments of intense X-ray sources at larger facilities are now making nonlinear X-ray measurements possible. X-ray transient grating spectroscopy[58] can directly give element-specific measurements of vibrational, magnetic, or electronic transport. XUV second harmonic generation anisotropy has measured atomic displacement with sub-angstrom spatial resolution.[59] X-ray sum frequency generation is allowing an element-specific addition to an already powerful technique for studying interfaces.[58] X-ray wave mixing processes are also starting to reach the level of complexity needed for true two-dimensional spectroscopy studies of coupled excitations.[60,61]

## Acknowledgements

This material is based upon work supported by the Air Force Office of Scientific Research under award number FA9550-21-1-0022. I.M.K. was supported by an NSF Graduate Research Fellowship (DGE-1745301). J.M.M. was supported by the Liquid Sunlight Alliance, which is supported by the U.S. Department of Energy, Office of Science, Office of Basic Energy Sciences, Fuels from Sunlight Hub under Award Number DE-SC0021266.